\documentclass[12pt,a4paper]{article}
 \pdfoutput=1
\usepackage[left=1cm,right=1cm,top=2cm,bottom=2cm]{geometry}
\usepackage{float}
\usepackage{graphicx}

\title{Static stability of collapsible tube \\ conveying non-Newtonian fluid}

\author{V.S. Yushutin  \\
  \multicolumn{1}{p{.8\textwidth}}{\centering\emph{Institute of Mechanics of Lomonosov Moscow State University, Moscow, Russia}}}

\begin{document}

\maketitle{}

\begin{abstract}
The global static stability of a Starling Resistor conveying non-Newtonian fluid is considered. The Starling Resistor consists of two rigid circular tubes and axisymmetric collapsible tube mounted between them. Upstream and downstream pressures are the boundary condition as well as external to the collapsible tube pressure. Quasi one-dimensional model has been proposed and a boundary value problem in terms of nondimensional parameters obtained. Nonuniqueness of the boundary value problem is regarded as static instability. The analytical condition of instability which defines a surface in parameter space has been studied numerically. The influence of fluid rheology on stability of collapsible tube  is established.
\end{abstract}  

\section{Introduction}

\begin{figure}[h!]
\center{\includegraphics[width=17cm]{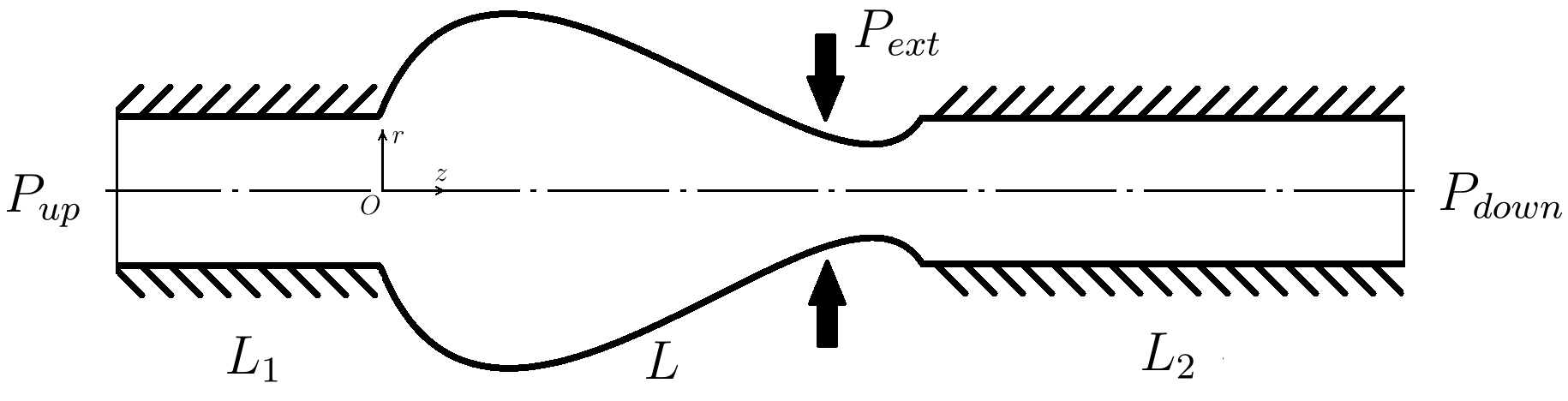}}
\caption{Mechanical model of a Starling Resistor}
\label{geometry}
\end{figure}

The present paper is aimed to investigate global static stability of a prestressed axisymmetric collapsible elastic tube conveying a non-Newtonian fluid. The mechanical model under consideration is an axisymmetric variant of plane Starling Resistor previously discussed in many studies: a flexible thin tube of finite length is mounted from both ends on rigid parts which are subjected to fixed upstream and downstream pressures (Fig. \ref{geometry}).

The problem has its origins mainly in studies of blood flow, air expiration from lungs and other biological situations when a media is forced to flow through a compliant channel. The problem has been investigated from different points of view,  the papers \cite{Heil_Jensen}-\cite{Grotberg_Jensen} are recommended for a review purpose.

The fluid is generally considered as Newtonian in the problem of media flow through a collapsible tube. But it's well known that most biofluids exhibit non-Newtonian properties which haven't been taken into account (to the author knowledge) in any paper concerning stability of the system. The influence of fluid rheology on local (TWF) stability has been investigated in \cite{Yushutin}. It has been shown how material nonlinearity affects the local properties of elastic wave-like disturbances.

It seems like we are still far from entire understanding of the physical phenomenon even in the case of linear Newtonian fluid flow through a collapsible tube but the paper is to shed light to the other aspect of the problem - fluid nonlinearity. So the study is devoted to developing of a model of power-law fluid flow past a flexible tube and examining its global static stability.

\subsection*{Power-law fluid}

The non-Newtonian fluid is assumed to have power-law (Ostwald - de Waele) properties.
The power-law fluid obeys the following constitutive relation:
$$\sigma_{ij}{}=-p{} \;I_{ij}+2\mu{}\,v_u^{n-1}v_{ij}\,,\eqno{(1)}$$
where $p$  is a pressure, $I_{ij}$ is the unit tensor, $v_u=\sqrt{v_{ij}v_{ij}}$ is the second invariant of strain rate $v_{ij}$.

There are two material constants: $\mu$ and $n$. The first is a dynamic viscosity and has the dimensions of $Pa\cdot{}s^n$ while the latter, $n$, is a nondimensional parameter which characterizes the physical nonlinearity of the fluid. In case of $n=1$ the media becomes of Newtonian type. Also the fluid is assumed to be incompressible and its density is denoted as $\rho$.

\section{Mathematical formulation}

A quasi one-dimensional model of a power-law fluid propagation through a collapsible tube will be proposed in the section.
Consider an axisymmetric flow past a channel whose boundary can be characterized by the radius $R(z,t)$ of circular cross-section, where $z$ is longitudinal coordinate and $t$ is the time. The velocity vector $(v_r,v_z,0)$  has two nonzero parts with respect to cylindrical coordinate system. One needs to average the fluid flow over a cross section to obtain a quasi one-dimensional model \cite{Formaggia}.
\subsection*{Karman-Pohlhausen approximation}
To average the mechanical system under investigation one should apply some prescribed velocity profile on a cross-section $S(z,t)$.
The following approximation (Karman-Pohlhausen) is assumed in this paper:
$$v_z(r,z,t)=\frac{Q(z,t)}{\pi{}R^2(z,,t)}s\left(\frac{r}{R(z,t)}\right)\,,\eqno{(2)}$$
where $Q(z,t) = \int_{S(z,t)}v_z(r,z,t)dS$ is the volume flux through the cross-section and $s(x) = \frac{3n+1}{n+1}\left(1-x^{(n+1)/n}\right)$ is a prescribed profile corresponding to Poiseuille solution.

 \subsection*{Long wavelength assumption}
The equations of motion and continuity can now be integrated over the arbitrary cross-section $S(z,t)$ applying a profile law (2). One can assume that $v_r/v_z\sim{}\partial{}R/\partial{}z\sim{}\varepsilon$ is a small parameter to obtain a one-dimensional model described only in terms of mean (with respect to cross section $S(z,t)$) parameters:

$$\frac{\partial{Q(z,t)}}{\partial{t}}+\frac{\partial{}}{\partial{z}}\left(\frac{\alpha(n) Q(z,t)^2}{2\pi{}R(z,t)^2}\right)+\frac{\mu{}}{\rho{}}\frac{K(n)
Q(z,t)^n}{\pi{}^{n-1}R(z,t)^{3n-1}}+
\frac{\pi{}R(z,t)^2}{\rho{}}\frac{\partial{P(z,t)}}{\partial{z}}=0\,.\eqno{(3)}$$
$$\frac{\partial{Q(z,t)}}{\partial{z}}+\frac{\partial{\left(\pi{}R^2(z,t)\right)}}{\partial{t}}=0\eqno{(4)}$$
$$\alpha(n)=\frac{2(3n+1)}{(2n+1)}\,,\quad{}K(n)=\frac{(3n+1)^n 2^{(3-n)/{2}}}{n^n}$$
Here $P(z,t)=\left(\int_{S(z,t)}p(r,z,t)dS\right)/\pi{}R^2(z,t)$ is the mean value of pressure. A description of tube mechanical behaviour is needed to obtain a closed system of equations.

 \subsection*{Massless string model}
A massless string model \cite{Quarteroni} can be used to describe the tube response to fluid pressure. This model neglects longitudinal displacement of tube due to fluid drag.
In contrast to plane Starling Resistor the tube cross-section provides an additional stiffness $\beta{}$ due to a circular shape:
$$P(z,t)-P_{ext}(z)=\beta{}\left(R(z,t)-R_0\right)-F\frac{\partial{}^2{R(z,t)}}{\partial{z^2}}\;,\;\;\beta{}=\frac{E_sh_s}{(1-\nu^2_s)R_0^2}\,,\eqno{(5)}$$
where $\beta{}$ - radial stiffness, $E_s$ and $ \nu{}_s$
are the Young's modulus and the Poisson's ratio of the tube's material, $R_0$ - undeformed radius of the tube, $h_s$ - the
thickness of the tube's wall, $P_{ext}(z)$ - an external pressure and $F$ - the force (per unit perimeter) of longitudinal tension.

\subsection*{Boundary conditions}
The prescribed upstream $p(-L_1)=P_{up}$ and downstream pressure $p(L+L_2)=P_{down}$ are the boundary conditions of the problem.
The equations (3) and (4) are valid through the whole system, while the equation (5) can be applied only to the compliant region, $z \in [0,L]$. Therefore, it's useful to solve the equations (3) and (4) in the upstream ($z \in [-L_1,0]$) and downstream ($z \in [L,L+L_2]$) rigid segments and hence to carry down pressure boundary conditions to an entrance ($z=0$) and an exit ($z=L$) of the collapsible tube \cite{Xu}.

\subsection*{Reference solution}
Generally the external pressure $P_{ext}$ is arbitrary but in the study it will be prescribed in a nontrivial way. Applying the same upstream pressure $P_{up}$ and the downstream pressure $P_{down}$ to the absolutely rigid Starling Resistor results in constant pressure gradient $k_0=(P_{up}-P_{down})/(L+L_1+L_2)$ and corresponding flux $q_0 = \left( k_0\pi{}^n R_0^{3n+1}/\mu{}K(n)\right)^{1/n}$ (Hagen-Poiseuille law). One can set $$P_{ext}(z) = P_{up} - k_0( z+L_1 ) $$ to balance the internal and external to the membrane forces and to have the uniform solution $R(z)=R_0$ even in the case of compliant properties of collapsible segment. So the special external pressure applied is a simplification of a general case which is made in order to deal only with uniform reference solution:
  $$R(z,t)=R_0\,,\quad{}Q(z,t)=q_0\,,\quad{}P(z,t)=P_{ext}(z)\,.$$
  This reference solution of system of equations (3)-(5) will be perturbed to explore static stability of the system in the next section.

\section{Static stability}

The statically perturbed solution corresponds to different volume flux $Q(z,t)= q$, tube shape $R(z,t)=r(z)$. The substitution of this perturbed solution to stationary form of equation (3)-(5) leads only to one equation:
$$\frac{d{}}{d{z}}\left(\frac{\alpha(n) q^2}{2\pi{}r(z)^2}\right)+\frac{\mu{}}{\rho{}}\frac{K(n)
q^n}{\pi{}^{n-1}r(z)^{3n-1}}+
\frac{\pi{}r(z)^2}{\rho{}}\left(\beta{}\frac{d{r(z)}}{d{z}}-F\frac{d^3r(z)}{dz^3}-k_0\right)=0\,.\eqno{(6)}$$

Consider a stationary flow with flux $q$ through the rigid upstream (downstream) segment. Equation~(3) lead to the well-known relation between the flux and a constant pressure gradient $k=-{\partial{P(z,t)}}/{\partial{z}}$:
$$k=\frac{q^n\mu{}K(n)}{\pi{}^n R_0^{3n+1}}\,.$$
Note that value $k$ retains the meaning of pressure gradient only in the rigid segment of Starling Resistor, and it allows to calculate pressure distribution over the length of the upstream (downstream) rigid segment:
$$p(z)=P_{up}-k(z+L_1)\quad{}\left(p(z)=P_{down}+k(L_2+L-z)\right)$$

Now the upstream and downstream pressure boundary conditions can be carried down to the collapsible segment ends in order to restrict mathematical region to the segment $[0,L]$. The entrance pressure $P_1=p(0)$ from the one hand can be expressed as $P_{up}-k L_1$ and from the other hand can be obtained from equation (5) giving the first boundary condition at $z=0$:
$$L_1(k-k_0)=F\frac{d^2r(0)}{dz^2}\,,\quad{}r(0)=R_0\,.\eqno{(7)}$$
The latter is a pinned boundary condition. Using the same approach one can find another set of conditions at the exit boundary:
$$L_2(k-k_0)=-F\frac{d^2r(L)}{dz^2},\,\quad{}r(L)=R_0\,.\eqno{(8)}$$

The system of equations (6)-(8) forms an nonlinear boundary value problem having $q$ as an eigenvalue and $r(z)$ as an eigenfunction. This boundary value problem will be investigated through all next sections.

\subsection*{Boundary value problem}

Let us introduce the nondimensional parameters:
$$n,\,\, \xi_1=\frac{L_1}{L},\,\xi_2=\frac{L_2}{L},\, S=\beta{}\frac{\rho{}^{\frac{n}{2-n}}R_0^{\frac{n+2}{2-n}}}{\mu{}^{\frac{2}{2-n}}}\varepsilon{}^{\frac{2}{2-n}}\,,$$$$ T=F\frac{\rho{}^{\frac{n}{2-n}}R_0^{\frac{3n-2}{2-n}}}{\mu{}^{\frac{2}{2-n}}}\varepsilon{}^{\frac{2(3-n)}{2-n}}\,,R=q_0\frac{\rho{}^{\frac{1}{2-n}}R_0^{\frac{3n-4}{2-n}}}{\pi{}\mu{}^{\frac{1}{2-n}}}\varepsilon{}^{\frac{1}{2-n}}\,,
\quad{}\eqno{(9)}$$
where $n$ is the power-law fluid nondimensional material constant, $\xi_i$ characterize the system geometry, $R$ is a modified Reynolds number, $T$ is the nondimensional longitudinal force  and $S$ is the nondimensional radial stiffness. The dimensional basis \{$\mu$, $\rho$, $R_0$\} is chosen to vary experimental parameters $q$, $F$, $S$ independently.

Considering only small shell displacements allows to linearize the boundary value problem with respect to a small parameter $a$:
$$r(z)=R_0\left(1+ay(z/L)\right),\quad q=q_0(1+a\lambda{}), \quad f(x)=y(x)/\lambda{}\,,\quad x=z/L,\lambda{}\neq{}0\,,$$
$$T {\frac {{d}^{3}f \left( x \right)}{{ d}{x}^{3}}}
  + \left( {R}^{2}\alpha \left( n \right) -S \right) {\frac
{df \left( x \right)}{{ d}x}} +K\left( n \right) {R}^{n}\, \left(  \left( 3n+1
 \right) f \left( x \right) -n \right) =0\,,\eqno{(10)}$$

$${\frac {{d}^{2}f \left( 0 \right)}{{d}{x}^{2}}} ={\frac {K
 \left( n \right) {R}^{n}n\xi_1}{T}}\,,\quad{}f(0)=0\,,\eqno{(11)}$$
$${\frac {{d}^{2}f \left( 1 \right)}{{d}{x}^{2}}} =-{\frac {K
 \left( n \right) {R}^{n}n\xi_2}{T}}\,,\quad{}f(1)=0\,.\eqno{(12)}$$

The linear boundary value problem (10)-(12) is of third order and has four boundary conditions. A condition of the problem solvability defines the neutral surface in nondimensional parameter space on which the nontrivial solution of the system (6)-(8) can be found.

The general solution of equation (10) can be found in a following complex form (except the case of multiple roots):
$$f(x) = \frac{n}{3n+1} + \sum_{i=1}^3 c_i e^{m_i x}\,,\eqno{(13)}$$
where the values $m_i$ are the roots of corresponding characteristic equation. Substituting (13) in (11), (12) gives a forth order linear system with respect to three complex unknown $c_i$. One can find an expression for a neutral surface by making the determinant of augmented matrix equal to zero.

 Numerical analysis of this expression in a space of nondimensional parameters $n$, $\xi_1$, $\xi_2$, $R$, $T$, $S$  will be conducted through the next sections.

\section{Numerical analysis of neutral surface}
Neutral surface consists of those points of nondimensional parameter space where nontrivial solution of (10)-(12) exists. Existence of nontrivial solution means that the reference solution can be statically unstable.

The term "neutral curve" applies to two-dimensional cross section of neutral surface when all the parameters are set except for two of them.

\subsection*{"Automodel" case: $S=\alpha{}(n) R^2$}
\begin{figure}[H]
\center{\includegraphics[width=17cm]{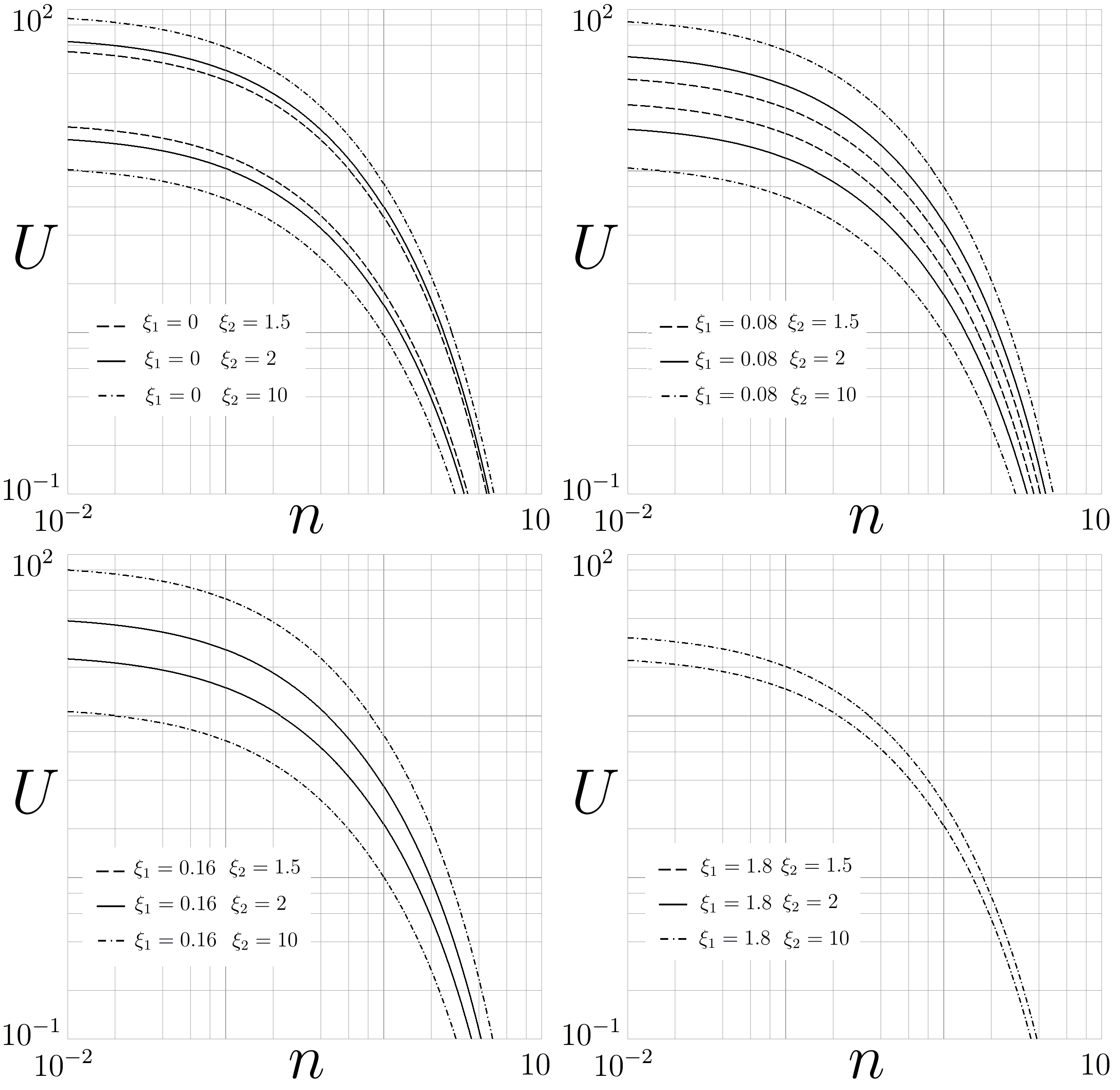}}
\caption{Neutral curves in the automodel case $S=\alpha{}(n) R^2$ in the space of $U={R^n}/{T}$ and $n$.}
\label{automodel}
\end{figure}

For a given values of nondimensional parameter $n$, $\xi_1$, $\xi_2$ there always exists a special case  $ S=\alpha{}(n) R^2$, when the boundary value problem (10)-(12) has an "automodel" solution. One can see that a variable $$U=\frac{R^n}{T}$$ governs the equation and the boundary conditions in this situation. Hence the expression for neutral surface in nondimensional parameter space can be reduced to single equation on $U$ having $n$ as parameter. Each solution $U_0$ defines a straight line in ($\log{T}$,~$\log{R}$)-subspace which is very useful to compare with as $S\to{}\alpha{}(n)R^2$. Relations between $U_0$ and $n$ are shown on Fig. \ref{automodel}.

Numerical investigation shows that there can be no solution $U_0$ for particular values of geometry parameters, for example: $\xi_1=0$ and $\xi_2=0$. This corresponds to a collapsible tube without rigid parts but with prescribed pressure on fixed ends. For any $n$ the automodel system is statically stable in this case since nontrivial solution $U_0$ of (10)-(12) can not be found.

Physically the "automodel" case corresponds to sub- to supercritical flow transition when small amplitude long-wavelength traveling wave flutter appears \cite{Yushutin} (the value $\chi = \alpha{(n)} R^2/S$ was used there to distinguish sub- from supercritical regime).

\subsection*{General case}
In this section the neutral surface will be plotted through two-dimensional cuts by (${T}$,~${R}$)-subspace for different values of stiffness $S$ and power-law index $n$ with other nondimensional parameters being set as follows:
$$\xi_1=0\,,\quad{} \xi_2=0\,.$$
For classical Penrose tubes parameters $S$ is of order of millions and for typical real fluids $0<n<2$.
It is also useful to distinguish sub- and supercritical regimes on this plots. The transition line $R=R_{tr}=\left(S/\alpha{}(n)\right)^{1/2}$ is horizontal and it  corresponds to automodel case discussed above. The region below the transition line ($R<R_{tr}$) is subcritical.

Note that for higher stiffness $S$ all the neutral curves move upwards. Smaller values ($n<1$) which corresponds to pseudoplastic behaviour of the fluid result in considerable defomation of neutral curves.

\begin{figure}[H]
\center{\includegraphics[width=17cm]{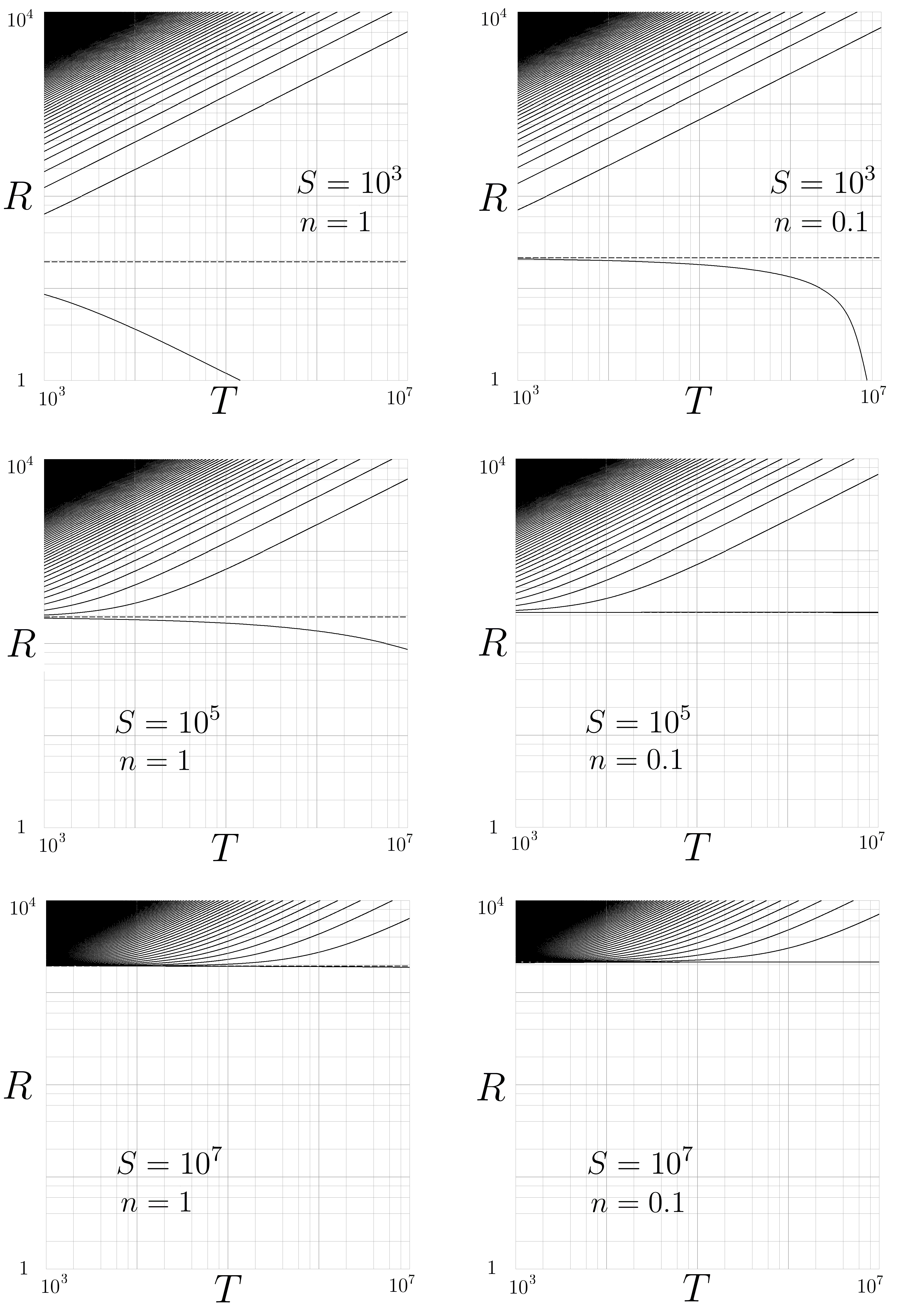}}
\caption{Neutral curves for different values of stiffness $S$ and power-law index $n$ in the case of $\xi_1=0$ and $\xi_2=0$. The dashed line corresponds to sub- to supercritical regime transition.}
\label{general}
\end{figure}

\section{Discussion and conclusions}

In this paper the global static stability of axisymmetric Starling Resistor conveying power-law fluid has been investigated. The global stability is found to be largely influenced by the physical nonlinearity $n$ of the fluid.

The external pressure was chosen to be of special form giving uniform reference solution. Mechanical response of the tube resembles generalised "string" model. The steady equations of motion and incompressibility lead to a boundary value problem which has a nontrivial solution only for specific values of nondimensional parameters. Thus the static instability is associated with nonuniqueness of boundary value problem solution.  The analytical condition of nonuniqueness is investigated numerically for different values of parameters including the automodel case.

The automodel case corresponds to transition from sub- to supercritical regime. It has been found that the transition isn't always statically unstable. The neutral curves has been plotted at Fig. \ref{automodel} for different geometric parameters $\xi_1$ and $\xi_2$. Increasing of $\xi_2$ generally leads to neutral curves appearance while sufficient growth of $\xi_1$ vanishes them. Also one can see how critical values of automodel parameter $U=R^n/T$ is affected by nonlinearity: the larger $n$ the lower the value $U_0$.

The transition between regimes is plotted as dashed line in (T,R)-space for different values of stiffness $S$ and power-law index $n$ (Fig. \ref{general}). One can see that static instability can occur for values of $R$ lower than $R_{tr}$ which corresponds to transition. It's worth to point out that increasing of stiffness $S$ move upwards all the neutral curves. Decreasing of $n$ makes the smallest value of $R$ larger when $T$ is fixed.

The problem under consideration is treated in very simplified matter. But since one-dimensional models are believed to capture qualitative properties of Starling Resistor it's clearly showed that the nonlinearity of the fluid greatly influences the stability condition.

\end{document}